\documentclass[aps,amssymb,amsmath,pra, twocolumn,superscriptaddress]{revtex4}
\usepackage{amsmath}
\usepackage{graphicx}
\usepackage{amsfonts}
\usepackage{epsfig}
\usepackage{hyperref}
\usepackage{amssymb}
\usepackage{bm}
\usepackage{bbm}
\usepackage{subfigure}
\newcommand{\be}{\begin{equation}}
\newcommand{\ee}{\end{equation}}
\newcommand{\bc}{\begin{center}}
\newcommand{\ec}{\end{center}}
\newcommand{\bea}{\begin{eqnarray}}
\newcommand{\eea}{\end{eqnarray}}

\begin{document}
\title{Disordered quantum walk-induced localization of a Bose-Einstein condensate}
\author{C. M. \surname{Chandrashekar}}
\email{chandru@imsc.res.in}
\affiliation{Center for Quantum Sciences, The Institute of Mathematical Sciences, Chennai 600113, India}

\begin{abstract}

We present an approach to induce localization of a Bose-Einstein condensate in a one-dimensional lattice under the influence of unitary quantum walk evolution using disordered quantum coin operation.
We introduce a discrete-time quantum walk model in which the interference effect is modified to diffuse or strongly localize the probability distribution of the particle by assigning a different set of coin parameters picked randomly for each step of the walk, respectively. Spatial localization of the particle/state is explained by comparing the variance of the probability distribution of the quantum walk in position space using disordered coin operation to that of the walk using an identical coin operation for each step. Due to the high degree of control over quantum coin operation and most of the system parameters, ultracold atoms in an optical lattice offer opportunities to implement a disordered quantum walk that is unitary and induces localization. Here we present a scheme to use a Bose-Einstein condensate that can be evolved to the superposition of its internal states in an optical lattice and control the dynamics of atoms to observe  localization. This approach can be adopted to any other physical system in which controlled disordered quantum walk can be implemented.
\end{abstract}

\maketitle
\preprint{Version}

\section{Introduction}
\label{intro} 
Localization of waves in disordered media originally was predicted in the context of transport of electrons in disordered crystals by Anderson \cite{And58}. Anderson localization,  resulting in the absence of diffusion, originates from the interference between multiple scattering paths (cf. \cite{LR85}). This phenomenon is now ubiquitous in physics \cite{KM93, Tig99}; it has been experimentally observed and theoretically studied in a variety of systems, including light waves \cite{AL85, WBL97, SLT99, SGA06, SBF07, LAP08} and matter waves \cite{DZS03, RDF08, BJZ08, Adh10}. 
\par
Quantum walks (QWs) \cite{Ria58, FH65, ADZ93, DM96, FG98}, which are the quantum analog  of the  classical random walks (CRWs), evolve in position space involving interference of amplitudes of multiple traversing paths. The quantum features of interference and superposition are know to make probability of QW spread quadratically faster with time than its classical counterpart in one- dimension. Some studies have shown the localization of the interference of amplitudes between multiple traversing paths of the QW distribution around the origin from various different perspectives \cite{RAS05, OKA05, YKE08, Kon10, Cha10a, SK10, CGB10}. In particular, it is shown that the localization of the walk dynamics in one dimension can be controlled by introducing drifts with constant momentum between two consecutive steps of QW \cite{RAS05} or by evolving the walk in a random medium characterized by a static disorder \cite{YKE08}. 
\par
The key factor for the interference effect to result in localization is the broken periodicity in the dynamics of the system induced by the disordered media. However, broken
periodicity need not be mediated by a disordered or a random medium alone; operations deÞning the dynamics of the system can be made random to break the periodicity such that they
mimic the effect of a random medium in the system and manifest localization. Taking this into consideration, we can carefully construct a QW evolution on a physical system such
that the dynamics of the walk without any disorder in the lattice is similar to the dynamics of a walk in a disordered medium or in the lattices leading to localization. Owing to the high degree of control over most of the system parameters and recent experimental implementation of QW  \cite{KFC09}, ultracold atoms in an optical lattice offer opportunities for the study of disordered QW-induced localization. 
\par
Using the ßexibility and control that has been achieved
over the ultracold atoms, Bose-Einstein Condensate (BEC) has
been used for the study of disorder-induced localization. Using
a cigar-shaped noninteracting BEC, exponential localization
 \cite{BJZ08} and a crossover between extended and exponentially
localized states \cite{RDF08} have been experimentally demonstrated.
Billy {\em et al.} \cite{BJZ08} demonstrated exponential localization of
$^{87}$Rb atoms when released into a one-dimensional waveguide
in the presence of a controlled disorder created by a laser
speckle. Roati {\em et al.} \cite{RDF08}, using $^{39}$K atoms, demonstrated its
localization in a one-dimensional bichromatic optical-lattice
potential created by the superposition of two standing-wave
polarized laser beams with different wavelengths. In the
preceding two experiments, localization was demonstrated
through investigations of the transport properties and spatial
and momentum distributions. Numerical study of localization
of BEC in a bichromatic optical-lattice potential \cite{AS09} and in a random potential \cite{CA10} has been reported. Recently, it was shown that the localization by bichromatic potentials is
produced by a trapping by the potential and is not due to a
quantum suppression, in contrast to the Anderson model \cite{AL10}.
There are also other time-dependent phenomena that can
induce population imbalance of two self-interacting BEC and,
hence, localization in one of the double-well potentials unlike
Anderson localization, which is a stationary phenomenon \cite{SFGS97}. 
\par
Taking the preceding points into consideration, in this
paper, we present a new scheme to observe dynamic localization of ultracold atoms in an optical lattice. Periodicity
in the dynamics of atoms in an optical lattice is broken
using a disordered evolution of the QW and this leads to the
interference of amplitude of multiple traversing paths of atoms
in an optical lattice to localize. Direct control over the quantum
coin operation makes it possible to choose the random set of
coin operations and control the dynamics of the QW  \cite{DM96, CSL08},
which, in turn, allows us to break the periodicity and localize
the evolution. In particular, we discuss localization using a
BEC with noninteracting and interacting atoms, respectively.
This scheme can be expanded to any of the physical systems
on which the QW can be implemented.
\par
This article is arranged as follows. In Sec.  \ref{qw}, we describe
the DTQW model on a line. In Sec.  \ref{ddtqw}, we introduce the
disordered DTQW model using controllable quantum coin
operation randomly picked for each step of walk and present
both diffusive and localization effects, respectively. In Sec.  \ref{LBEC},
we discuss the implementation of QW in ultracold atoms and
localization of BEC. Finally, we conclude in Sec. \ref{conc}.

\section{Quantum walk}
\label{qw}

Like  their  classical counterparts,  QWs  are  also   widely  studied  in   two  forms, continuous-time  QW  (CTQW) \cite{FG98}  and  discrete-time QW  (DTQW)
\cite{ADZ93, DM96, ABN01, NV01}, and are found to be very useful from the perspective
of quantum algorithms \cite{Amb03, CCD03, SKB03, AKR05}. 
Furthermore, they have been  used to demonstrate the coherent  quantum control over
atoms \cite{CL08} and to explain phenomena such
as the breakdown of an electric-field driven  system \cite{OKA05} and wavelike energy  transfer  within
photosynthetic systems \cite{ECR07,  MRL08}. Generation of entanglement between two spatially separated systems is another application of QWs \cite{CGB10}.  Experimental implementation of  QWs has  been reported with  samples in nuclear magnetic resonance (NMR) systems \cite{DLX03, RLB05}, in the continuous
tunneling of light fields  through waveguide lattices \cite{PLP08}, in
the  phase space  of trapped  ions  \cite{SMS09, ZKG10}  based on  the
scheme proposed by \cite{TM02}, with single optically trapped atoms  \cite{KFC09}, and  with  single photons  \cite{SCP10, BFL10}.
Recently, implementation  of a QW-based  search algorithm in a NMR system
has  been  reported \cite{LZZ10}.   Various  other  schemes have  been
proposed for  their physical  realization in different  physical systems
\cite{RKB02, EMB05, Cha06, MBD06}.
\par
DTQW, the dynamics f which can be controlled by controlling the quantum coin operation, is used for the study presented in this paper. It  is  modeled  as  a  particle consisting  of a two-level coin  (a qubit) in the  Hilbert space  ${\cal H}_c$,
spanned  by $|0\rangle$  and  $|1\rangle$, and  a  position degree  of
freedom in  the   Hilbert  space  ${\cal  H}_p$,  spanned  by
$|\psi_x\rangle$, where $x \in {\mathbbm  I}$, the set of integers.
A  $t$-step DTQW with unit time required for each step of walk is generated by  iteratively 
applying  a unitary  operation $W$,  which acts  on the Hilbert space 
${\cal H}_c\otimes    {\cal     H}_p$:    
\be
|\Psi_t\rangle=W^t|\Psi_{in}\rangle,
\ee  
where 
\be
\label{inst}
|\Psi_{in}\rangle  = (\cos(\delta        /2)|0\rangle+       \sin(\delta/2)       e^{i\phi}
|1\rangle)\otimes |\psi_0\rangle
\ee
is an arbitrary initial state of the particle at the origin and $W\equiv S(B \otimes  {\mathbbm 1})$, where 
\be
B = B_{\xi,\theta,\zeta}       \equiv      \left(      \begin{array}{clcr}
  \mbox{~}e^{i\xi}\cos(\theta)      &     &     e^{i\zeta}\sin(\theta)
  \\ e^{-i\zeta} \sin(\theta) & & - e^{-i\xi}\cos(\theta) 
\end{array} \right)\in U(2)
\ee
is  the quantum coin operation.   $S$ is  controlled-shift operation          
 \be            S\equiv         \sum_x \left [  |0\rangle\langle
0|\otimes|\psi_x-1\rangle\langle   \psi_x|   +  |1\rangle\langle
1|\otimes |\psi_x+1\rangle\langle \psi_x| \right ]. 
 \ee
The probability to find the particle at site $x$ after $t$ steps is given by $p(x,t)  = \langle
\psi_x|{\rm tr}_c (|\Psi_t\rangle\langle\Psi_t|)|\psi_x\rangle$.

\begin{figure}[ht]
\begin{center}
\epsfig{figure=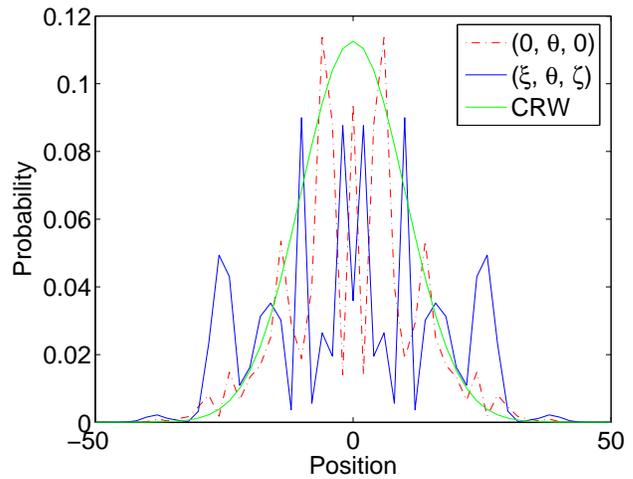, width=9.0cm}
\caption{(Color online) Probability distribution of the QW evolved by assigning quantum coin operation in a U(2) group with random parameters to each step of the walk.  Parameters are assigned from $\xi, \theta, \zeta \in \{0, \pi/2 \}$ for $B_{\xi, \theta, \zeta}$ and $B_{0, \theta, 0}$ respectively. The distribution is after 100 steps of the walk on a particle with initial state at the origin $|\Psi_{ins} \rangle  = \frac{1}{\sqrt 2}(|0\rangle + i |1\rangle) \otimes |\psi_{0}\rangle$. We see that the variance of the distribution is very much close to the variance of the classical random walk (CRW) distribution.} 
\label{fig:randomqw} 
\end{center}
\end{figure}

\section{Disordered discrete-time quantum walk}
\label{ddtqw}

Direct control over the quantum coin operation $B$ makes it possible to control the dynamics of the DTQW \cite{DM96, CSL08}. For a walk evolution on a particle in one dimension with initial state 
\be
|\Psi_{ins} \rangle  = \frac{1}{\sqrt 2}(|0\rangle + i |1\rangle) \otimes |\psi_{0}\rangle,
\ee
at the origin using an unbiased coin operation, that is, $B_{\xi,\theta,\zeta}$ with $\xi = \zeta = 0$, the variance after $t$ steps of walk is $[1 - \sin(\theta)] t^2$ \cite{CSL08} and a symmetric probability distribution in position space is obtained \cite{NV01, TFM03}. Nonzero values for parameter $\xi$ and $\zeta$ when $\xi \neq \zeta$ introduces asymmetry to the probability distribution  \cite{CSL08}. It is also shown that DTQW with $\delta \neq \pi/2$ in Eq. (\ref{inst}) returns asymmetric probability distribution, and parameters $\xi$ and $\zeta$ can be adjusted to make the distribution symmetric.
\par
In a standard DTQW evolution, identical coin operation is used during each step of the walk making it a periodic evolution. This order can be broken by choosing a different coin operation for each step of the walk or by choosing a different coin operation at each position space. For simplicity, we will consider a different coin operation during each step. Evolution of DTQW using disordered coin operation can be constructed by randomly choosing a quantum coin operator for each step from a  set of operators in the U(2) group.
\begin{figure}[ht]
\begin{center}
\epsfig{figure=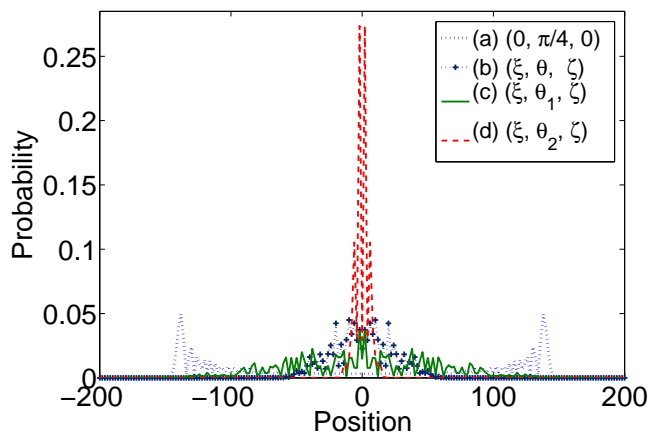, width=9.0cm}
\caption{(Color online) Comparison of the probability distribution of the QW evolved by assigning quantum coin operation in a U(2) group with random parameters in a different range to each step of the walk to that of the Hadamard walk.  (a) Hadamard walk, which is the standard form of the DTQW with identical coin ($\theta = \pi/4$), (b) the walk using  parameters assigned from $\xi, \theta, \zeta \in \{0, \pi/2 \}$ for $B_{\xi, \theta, \zeta}$, (c) the walk using  parameters assigned from $\xi, \zeta \in \{0, \pi/2 \}$ and $\theta_{1} \in \{0, \pi/4 \}$ and  (d) the walk using  parameters assigned from $\xi, \zeta \in \{0, \pi/2 \}$ and $\theta_{2} \in \{\pi/4, \pi/2 \}$.  Localization is seen in the case of (d). The distribution is after 200 steps of the walk on a particle with initial state at the origin $|\Psi_{ins} \rangle  = \frac{1}{\sqrt 2}(|0\rangle + i |1\rangle) \otimes |\psi_{0}\rangle$.} 
\label{fig:randomqw1} 
\end{center}
\end{figure}
That is,
\be
S(B_{\xi_{t}, \theta_{t}, \zeta_{t}}\otimes {\mathbbm
1})    \cdots S(B_{\xi_{x}, \theta_{x}, \zeta_{x}}\otimes {\mathbbm
1}) \cdots  S(B_{\xi_{0}, \theta_{0}, \zeta_{0}}\otimes {\mathbbm
1}) |\Psi_{in}\rangle
\ee
with randomly chosen parameters $\xi, \theta, \zeta \in \{0, \pi/2\}$ for each step. Though the coin parameters are randomly chosen for each step, the evolution is unitary and involves interference of amplitudes, and the effect is seen in probability distribution, see Fig. \ref{fig:randomqw}. 
\par
From the numerical evolution of 100 step QW shown in Fig. \ref{fig:randomqw}, although interference effect is seen, we note that the variance of the distribution is much closer to the variance of the CRW distribution. However, by restricting the range of the coin parameters that can be used for the walk, the probability distribution can be localized or made to diffuse in position space. One simple example we can consider is to by randomly pick different $\theta$ for each step from a subset of a complete range of $\theta$, with subsets $\theta_{1} \in \{0, \pi/4\}$ and $\theta_{2} \in \{\pi/4, \pi/2\}$, for $B_{\xi, \theta_{1}, \zeta}$ and $B_{\xi, \theta_{2}, \zeta}$ respectively, while other parameters are still picked from the complete range, $\xi, \zeta \in \{0, \pi/2\}$. The probability distribution obtained is shown in Fig. \ref{fig:randomqw1}. The QW using $\theta_{1} \in \{0, \pi/4\}$ diffuse in position space without any sharp peaks when compared to the standard QW evolution using identical coin operation  with two sharp peaks. The walk using $\theta_{2} \in \{\pi/4, \pi/2\}$, however,  localizes the distribution around the origin \cite{Cha10a}. For the numerical evolution, the parameter $\theta$ for each random coin operation in the evolution was generated from a random number generator program with an equal probable appearance of any number in the specific range. 

\section{Localization of Bose-Einstein Condensates}
\label{LBEC}
Degenerate atomic gases have been used as a system to experimentally implement a number of basic models in condensed matter theory. The possibility to create both, ordered and disordered lattice potentials in higher physical dimensions, the control of interatomic interactions, and the possibility to measure atomic density profile via direct imaging are the key advantages of atomic quantum gases (cf. \cite{LSA07, BDZ08}). 
\par
\begin{figure}[ht]
\begin{center}
\epsfig{figure=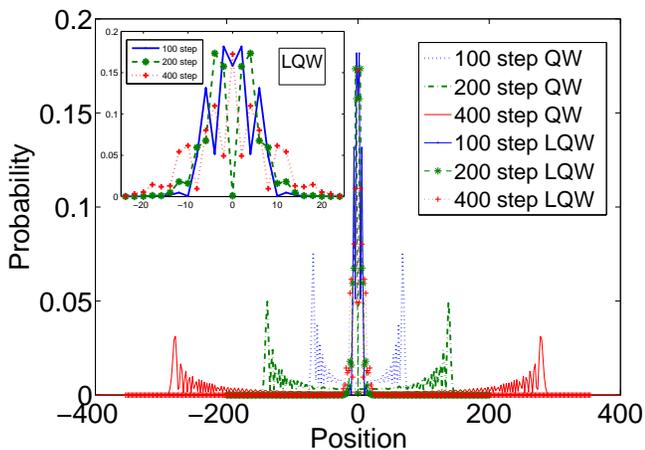, width=9.0cm}
\caption{(Color online) Comparison of a standard QW evolution (Hadamard walk) and a localized QW (LQW) after 100, 200, and 400 steps respectively.  For a walk using randomly picked parameters 
 $\theta_{2} \in \{\pi/4, \pi/2\}$ for each step of the walk, distribution remains localized near the origin
 irrespective of the number of steps (time ). The distribution is after the walk on a particle with initial state at the origin $|\Psi_{ins} \rangle  = \frac{1}{\sqrt 2}(|0\rangle + i |1\rangle )\otimes |\psi_{0}\rangle$. Inset is a closer view of localized QW distribution between position -30 and +30 for 100, 200 and 400 steps, respectively.} 
\label{localizedqw} 
\end{center}
\end{figure}
\begin{figure}[ht]
\begin{center}
\epsfig{figure=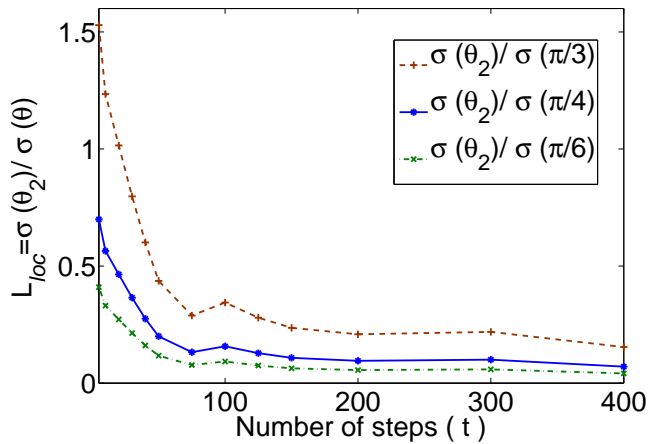, width=9.0cm}
\caption{(Color online) Localization length obtained by calculating  the ratio of the variance of the probability distribution LQW to that of the QW using an identical coin operation for each step. For each plot different $\theta$ in the coin operation is picked for an identical coin operation.} 
\label{fig:loclength} 
\end{center}
\end{figure}
Recently,  atomic gases were used to test and implement various quantum information protocols; particularly, by using a single atom in an optical lattice, experimental implementation of DTQW has been reported  \cite{KFC09}. Single laser cooled caesium (Cs) neutral atoms were deterministically delocalized over the sites of a one-dimensional spin-dependent optical lattice. Initially, the atoms distributed among the axial vibrational state was prepared in the $|0 \rangle \equiv |F = 4, m_{F} = 4\rangle$  hyperfine state by optical pumping, where $F$ is the total angular momentum, and $m_{F}$ is its projection onto the quantization axis along the dipole trap axis. The resonant microwave radiation, a $\pi/2$ pulse, was used to coherently couples this state to the $|1 \rangle \equiv |F = 3, m_{F} = 3\rangle$ state and initialize the system in the superposition $(|0\rangle + i|1\rangle)/ \sqrt{2}\otimes |\psi_{0}\rangle$.  The state-dependent shift operation is performed by
continuous control of the trap polarization, moving the spin state $|0\rangle$  to the right and state $|1\rangle$ to the left adiabatically along the lattice axis. After $t$ steps of coin operation and state-dependent shift, the final atom distribution is probed by fluorescence imaging. From these images, the exact lattice site of the atom after the walk is extracted and compared to the initial position of the atom. The final probability distribution to find an atom at position $x$ after $t$ steps  is obtained from the distance each atom has walked by taking the ensemble average over several hundreds of identical realizations of the sequence.  
\par
Experimental complexity aside, the preceding protocol can be adopted to other species of ultracold atoms and the non-interacting BEC as an initial state.  For example, rubidium ($^{87}$Rb) atoms in an optical trap with 
state $|F = 1, m_{F} = 1\rangle \equiv  |0 \rangle$ and $|F = 2, m_{F} = 2\rangle \equiv |1 \rangle$ can be coherently coupled and moved in position space to implement DTQW.  Implementing DTQW on the BEC involves evolving BEC into macroscopic superposition (Schr\"odinger cat) state \cite{LG85, MAK97, CLM98, DD02} after each shift operation which delocalizes the state in position space \cite{Cha06}. Evolving the BEC into a macroscopic superposition state to implement DTQW involves applying a rf pulse to the system to transfer the atoms part of the way between states $|0\rangle$ and $|1\rangle$. The duration of the pulse is kept much shorter than the self-dynamics of the condensate. This will only evolve each atom into the superposition of state $|0\rangle$ and $|1\rangle$, and the corresponding $N$ atom quantum state is a product of single-particle superposition, that is, it is still a microscopic superposition. However, as this initial state evolves under the nonlinear Hamiltonian that governs the BEC with attractive interparticle interactions, it reaches a macroscopic superposition in which all atoms in a given lattice are simultaneously in level $|0\rangle$ and $|1\rangle$,  $|\Psi_{BEC}\rangle = (1/\sqrt{2})[|N_{|0\rangle}, 0_{|1\rangle} \rangle + |0_{|0\rangle}, N_{|1\rangle} \rangle]$ \cite{DD02, Cha06}. 
Therefore, interatomic interactions are important to evolve the BEC into a macroscopic superposition state as discussed above. A DTQW on a noninteracting BEC as an initial state will implement the walk at an individual atom level and lose the coherence. We should note that the two experimental demonstration of localization \cite{BJZ08, RDF08} were done on a noninteracting BEC.  
\par
In a scheme proposed to implement a DTQW on ultracold
atoms in a BEC state \cite{Cha06}, it is Þrst evolved to a macroscopic
superposition state and a {\em stimulated Raman kick}, i.e., two
selected levels of the atoms are coupled to the two modes
of counterpropagating laser beams to coherently impart a
translation of atoms in the position space. After each translation, the wave packet is again evolved into the macroscopic
superposition state at each lattice $x$, where the number of
atoms $n_x < N$ and the process is iterated to implement a large
number of steps of QW. When $n_x$ is very small, the atoms will
evolve to the superposition only at an individual atom level,
i.e., microscopic superposition. With a certain modiÞcation to
this scheme, that is, by evolving atoms to the superposition
of the states at an individual atom level and implementing the
shift operation before the interatomic interaction takes over to
form a macroscopic superposition, the QW at an individual
atom level can be realized \cite{Cha10a}. 
\par
To observe localization, the periodic evolution using a $\pi/2$ pulse as quantum coin operation during each step of the walk in the preceding protocols is broken by randomly picking the pulse from the range $\{ \pi/4, \pi/2 \}$.  Therefore, during each step, the coin operation evolves the state to an unequal superposition state such that the constructive interference effect and amplitude are directed toward the origin.  
In Fig. \ref{localizedqw}, we compare the QW evolution using an identical coin operation for each step (Hadamard walk) and localized QW (LQW) after 100, 200, and 400 steps respectively.  For a walk using randomly picked parameters   $\theta_{2} \in \{\pi/4, \pi/2\}$ for each step of the walk, distribution remains localized near the origin irrespective of the number of steps ( time ).

In Fig. \ref{fig:loclength}, the  localization length $L_{\it loc}$ is numerically obtained by taking the ratio of the variance of the probability distribution LQW to that of the QW using identical coin operation for each step,
\be
L_{\it loc} = \frac{\sigma (\theta_2)}{\sigma (\theta)}
\ee
where $\theta_{2}$ is the randomly picked quantum coin parameter from range $\{ \pi/4, \pi/2 \}$ for each step and $\theta$ is the identical quantum coin parameter throughout the walk evolution.\\
\section{Conclusion}
\label{conc}
We have discussed an approach to dynamic localization
of ultracold atoms in a one-dimensional lattice under the
inßuence of DTQWusing disordered quantum coin operations.
We introduced a DTQW model in which a random coin
parameter is assigned to each step of the walk to break the
periodicity during the walk evolution. By picking the coin
operation from a different range of parameters, we have shown
that the DTQW on a two-state particle in a one-dimensional
lattice can be diffused or strongly localized in position space,
respectively. We have shown that these behaviors of the DTQW
can be efÞciently induced without introducing decoherence
into the system. Using ultracold atoms in an optical lattice
as a physical system, we have discussed implementation of a
DTQW ßexibility in control over the experimental parameters
to conÞgure walk to diffuse or localize in position space.
We have discussed implementation of DTQW on atoms at
an individual level in a BEC and on a BEC retaining the
macroscopic coherence (BEC) state throughout the evolution.
From this, we can conclude that the localization can be
observed in a BEC with noninteracting atoms and interacting
atoms, respectively. In general, the disordered coin operations
(microwave pulses) in DTQW on atoms can be made to
mimic the random media localizing the BEC. This approach
can be adopted to any other physical system in which a
controlled disordered DTQW can be implemented, broadening
the spectrum of possible application of DTQW to study
dynamics and phases in physical systems.
\\
\\
{\bf Acknowledgement :} The author would like to thank R. Simon for encouragement, and Subhashish Banerjee and Sandeep Goyal for stimulating conversations. 


\begin{thebibliography}{99}

\bibitem{And58} P. W. Anderson, Phys. Rev. {\bf 109}, 1492 (1958).

\bibitem{LR85} P. A.  Lee and T. V.  Ramakrishnan,  Rev. Mod. Phys. {\bf 57}, 287 (1985).

\bibitem{KM93} B. Kramer, A. MacKinnon,  Rep. Prog. Phys. {\bf 56}, 1469 (1993).

\bibitem{Tig99} V.  Tiggelen,  in {\it Wave Diffusion in Complex Media}, edited by J. P. Fouque (Kluwer, Dordrecht, 1999), p. 1.

\bibitem{AL85} M. P.  Van Albada and A. Lagendijk, Phys. Rev. Lett. {\bf 55}, 2692 (1985).

\bibitem{WBL97} D. S. Wiersma, P. Bartolini, A. Lagendijk, and R. Righini,  Nature {\bf 390}, 671 (1997).

\bibitem{SLT99} F. Scheffold, R. Lenke, R. Tweer, and G. Maret, Nature {\bf 398}, 206 (1999).

\bibitem{SGA06}  M. St\"orzer, P. Gross, C. M. Aegerter, and G. Maret, Phys. Rev. Lett. {\bf 96}, 063904 (2006).

\bibitem{SBF07} T. Schwartz, G. Bartal, S. Fishman, and M. Segev, Nature {\bf 446}, 52 (2007).

\bibitem{LAP08} Y. Lahini, A. Avidan, F. Pozzi, M. Sorel, R. Morandotti, D. N. Christodoulides, and Y. Silberberg, Phys. Rev. Lett. {\bf 100}, 013906 (2008).

\bibitem{DZS03}  B. Damski, J. Zakrzewski, L. Santos, P. Zoller, and M. Lewenstein,  Phys. Rev. Lett. {\bf 91}, 080403 (2003).

\bibitem{BJZ08} J. Billy, V. Josse, Z. Zuo, A. Bernard, B. Hambrecht, P. Lugan, D. Cle«ment, L. Sanchez-Palencia, P. Bouyer, and  A. Aspect, Nature {\bf 453}, 891 (12 June 2008).

\bibitem{RDF08}  G. Roati, C. DÕErrico, L. Fallani, M. Fattori, Chiara Fort, Matteo Zaccanti, Giovanni Modugno, Michele Modugno, and Massimo Inguscio, Nature {\bf 453}, 895 (12 June 2008). 

\bibitem{Adh10} S. K. Adhikari, Phys. Rev. A, {\bf 81} 043636 (2010).

\bibitem{Ria58}  G. V. Riazanov, Sov. Phys. JETP {\bf 6} 1107 (1958).

\bibitem{FH65} R. P. Feynman and A.R. Hibbs, {\it Quantum Mechanics and Path Integrals} (McGraw-Hill, New York, 1965).

\bibitem {ADZ93} Y. Aharonov, L. Davidovich and N. Zagury, Phys. Rev. A {\bf 48}, 1687, (1993).

\bibitem{DM96} D.  A. Meyer,  J. Stat. Phys. {\bf 85},  551 (1996).

\bibitem{FG98} E. Farhi and S. Gutmann, Phys.Rev. A {\bf 58}, 915 (1998).

\bibitem{RAS05} A. Romanelli, A. Auyuanet, R. Siri, G. Abal, and R. Donangelo, Physica A {\bf 352}, 409 (2005). 

\bibitem{OKA05} T. Oka, N. Konno, R. Arita, and H. Aoki, Phys. Rev. Lett. {\bf 94}, 100602 (2005). 

\bibitem{YKE08} Yue Yin, D. E. Katsanos, and S. N. Evangelou, Phys. Rev. A {\bf 77} 022302 (2008). 

\bibitem{Kon10} N. Konno, Quantum Information Processing, Vol.9, No.3, 405 (2010).

\bibitem{Cha10a} C. M. Chandrashekar, {\it Discrete-Time Quantum Walk - Dynamics and Applications}, arXiv:1001.5326 (2010). 

\bibitem{SK10} Y. Shikano and H. Katsura, arXiv:1004.5394 (2010).

\bibitem{CGB10} C. M. Chandrashekar, Sandeep K Goyal, and Subhashish Banerjee, arXiv:1005.3785 (2010).

\bibitem{KFC09} K. Karski, L. Foster, J.-M. Choi, A. Steffen, W. Alt, D. Meschede, and A. Widera, Science  {\bf 325}, 174 (2009).

\bibitem{AS09}S. K. Adhikari and L. Salasnich, Phys. Rev. A {\bf 80}, 023606 (2009).

\bibitem{CA10} Yongshan Cheng and S. K. Adhikari, Phys. Rev. A {\bf 82}, 013631 (2010). 

\bibitem{AL10} Mathias Albert and Patricio Leboeuf, Phys. Rev. A {\bf 81}, 013614 (2010).

\bibitem{SFGS97} A. Smerzi, S. Fantoni, S. Giovanazzi, and S. R. Shenoy, Phys. Rev. Lett. {\bf 79}, 4950 (1997).

\bibitem{CSL08} C. M. Chandrashekar, R. Srikanth, and R. Laflamme,  Phys. Rev. A {\bf 77}, 032326 (2008).

\bibitem{ABN01} A. Ambainis, E. Bach, A. Nayak, A. Vishwanath and J. Watrous, {\it Proceeding of the 33rd ACM Symposium on Theory of Computing} (ACM Press, New York, 2001), p.60.

\bibitem{NV01} A. Nayak and A. Vishwanath, DIMACS Technical Report, No. 2000-43  (2001) ; arXiv:quant-ph/0010117.

\bibitem{Amb03} A. Ambainis, Int. Journal of Quantum Information, {\bf 1}, No.4, 507-518 (2003).

\bibitem{CCD03} A. M. Childs, R. Cleve, E. Deotto, E. Farhi, S. Gutmann and D. A. Spielman, in {\it Proceedings of the 35th ACM Symposium on Theory of Computing} (ACM Press, New York, 2003), p.59.

\bibitem{SKB03} N. Shenvi, J. Kempe and K. Birgitta Whaley, Phys. Rev. A {\bf 67}, 052307, (2003).

\bibitem{AKR05} A. Ambainis, J. Kempe, and A. Rivosh, {\it Proceedings of ACM-SIAM
Symp. on Discrete Algorithms (SODA)},  (AMC Press, New York, 2005), pp.1099-1108.

\bibitem{CL08} C. M. Chandrashekar and R. Laflamme,  Phys. Rev. A  {\bf 78},  022314 (2008).

\bibitem{ECR07} G. S. Engel {\it et. al.}, Nature {\bf 446}, 782-786 (2007).  

\bibitem{MRL08} M. Mohseni, P. Rebentrost, S. Lloyd, A. Aspuru-Guzik, J. Chem. Phys. {\bf 129}, 174106 (2008).

\bibitem{DLX03} J. Du, H. Li, X. Xu, M. Shi, J. Wu, X. Zhou, and R. Han, Phys. Rev. A {\bf 67}, 042316 (2003)

\bibitem{RLB05} C. A.  Ryan, M.  Laforest, J. C. Boileau, and R. Laflamme, Phys. Rev. A
{\bf 72}, 062317 (2005).

\bibitem{PLP08} H. B. Perets, Y. Lahini, F. Pozzi, M. Sorel, R. Morandotti, and Y. Silberberg,  Phys. Rev. Lett. {\bf 100}, 170506 (2008).

\bibitem{SMS09} H. Schmitz, R. Matjeschk, Ch. Schneider, J. Glueckert, M. Enderlein, T. Huber, and T. Schaetz, Phys. Rev. Lett. {\bf 103}, 090504 (2009).

\bibitem{ZKG10} F. Zahringer, G. Kirchmair, R. Gerritsma, E. Solano, R. Blatt, and C. F. Roos, Phys. Rev. Lett. {\bf 104}, 100503 (2010) 

\bibitem{TM02} B. C. Travaglione and G. J. Milburn, Phys. Rev. A {\bf 65}, 032310 (2002).

\bibitem{SCP10} A. Schreiber, K. N. Cassemiro, V. Potocek, A. Gabris, P. Mosley, E. Andersson, I. Jex, and Ch. Silberhorn, Phys. Rev. Lett., {\bf 104}, 050502 (2010).

\bibitem{BFL10}  M. A. Broome, A. Fedrizzi, B. P. Lanyon, I. Kassal, A. Aspuru-Guzik, and A. G. White. Phys. Rev. Lett. {\bf 104}, 153602 (2010).

\bibitem{LZZ10} D. Lu, J. Zhu, P. Zou, X. Peng, Y. Yu, S. Zhang, Q. Chen, and J. Du, Phys. Rev. A {\bf 81}, 022308 (2010).

\bibitem{RKB02} W. Dur, R. Raussendorf, V. M. Kendon, and H. J. Briegel,  Phys. Rev. A {\bf 66}, 052319  (2002).

\bibitem{EMB05} K. Eckert, J. Mompart, G. Birkl, and M. Lewenstein,  Phys. Rev. A {\bf 72}, 012327 (2005).

\bibitem{Cha06} C. M. Chandrashekar,  Phys. Rev. A {\bf 74}, 032307 (2006).

\bibitem{MBD06} Z.-Y. Ma, K. Burnett, M. B. d'Arcy, and S. A. Gardiner, Phys. Rev. A {\bf 73}, 013401 (2006).

\bibitem{TFM03} B. Tregenna, W. Flanagam, R. Maile, and V. Kendon, New Journal of Physics {\bf 5}, 83 (2003). 

\bibitem{LSA07} M. Lewenstein,  A. Sanpera, V. Ahufinger, B. Damski,  A. Sen, and Ujjwal Sen, Adv. Phys. {\bf 56}, 243 (2007).

\bibitem{BDZ08} I. Bloch, J. Dalibard, and W. Zwerger, Rev. Mod. Phys. {\bf 80}, 885 (2008).

\bibitem{LG85} A.J. Leggett and Anupam Garg, Phys. Rev. Lett. {\bf 54} 857 (1985).

\bibitem{MAK97} M.-O. Mewes, M.R. Andrews, D.M. Kurn, D.S. Durfee, C.G. Townsend and W. Ketterle, Phys. Rev. Lett.{\bf 78}, 582-585 (1997).

\bibitem {CLM98} J.I. Cirac, M. Lewenstein, K. Momer and P. Zoller, Phys. Rev. A {\bf 57}, 1208, (1998).

\bibitem {DD02} D. A. R. Dalvit, and J. Dziarmaga , Los Alamos Sciences {\bf 27}, 166, (2002).

\end{thebibliography}
\end{document}